\documentclass[12pt]{article}

\newcommand{\w}{\mbox{\tiny $\wedge$}}

\date{}

\newcommand{\ww}{\mbox{\tiny $\wedge$}}

\title{Three-dimensional quantum geometry and black holes}

\author{M\'aximo Ba\~nados \\ 
Departamento de F\'{\i}sica Te\'orica, Universidad de
 Zaragoza, \\ Ciudad Universitaria 50009, Zaragoza, Spain.}

\begin{document}

\maketitle

\begin{abstract}
We review some aspects of three-dimensional quantum gravity with
emphasis in the `CFT $\rightarrow$ Geometry' map that follows from the
Brown-Henneaux conformal algebra.  The general solution to the
classical equations of motion with anti-de Sitter boundary conditions
is displayed.  This solution is parametrized by two functions which
become Virasoro operators after quantisation. A map from the space of
states to the space of classical solutions is exhibited. Some recent
proposals to understand the Bekenstein-Hawking entropy are reviewed in
this context. The origin of the boundary degrees of freedom arising in
2+1 gravity is analysed in detail using a Hamiltonian Chern-Simons
formalism.

%Invited talk at the Second Meeting ``Trends in Theoretical Physics",
%held in Buenos Aires, December, 1998.     

\end{abstract}

%------------------------------------
%------------------------------------
\section{Introduction}

General relativity is a highly complicated non-linear field theory
both classically and quantum mechanically. Even though a large number
of classical solutions exists, a general classification of the space
of solutions has never been achieved. The non-renormabizibility of
quantum gravity is not related to this issue, but if the general
structure of the space of solutions of the Einstein equations was
known, then the quantum version of phase space perhaps would be more
manageable. 

It is this aspect of three-dimensional gravity that makes it
attractive because the general solution to the equations of motion can
be written down. In this paper, we shall exploit this fact trying to
formulate a quantum theory of black holes by quantizing the space of
solutions directly.

Another important aspect of three-dimensional gravity is its
formulation as a Chern-Simons theory\cite{Achucarro-T}. Quantum
Chern-Simons theory is well understood for compact
groups\cite{Witten89,Moore-S,Labastida-R}. However, we shall be
interested in Euclidean
gravity with a negative cosmological constant whose associated group
is $SL(2,C)$, which is not compact. The quantization is then not
straightforward. We shall follow an alternative route by first solving
the equations of motion with prescribed boundary conditions and then
quantise. We shall see that the boundary conditions will play an
important role in making the quantum theory well-defined.

%--------------------------------------------
\subsection{Brief description of the results contained in this
article}
\label{Brief}

Let us start by briefly mentioning, without proofs, the main results
which will be of interest for us here. The relevant proofs will be
given below. We should remark at this point that most of the results
presented here are known in the literature in various contexts (the
relevant quotations will be given in the main text). The aim of
this article is to put things together in a self-contained
framework, and to explore some aspects of quantum black holes in
three dimensions. 

Let $M$ be a three dimensional manifold with a boundary denoted by
$\partial M$.  We assume that $\partial M$ has the topology of a
2-torus. Let $\{w, \bar w, \rho\}$ coordinates on $M$ such
that the boundary is located at $e^\rho =:l r \rightarrow\infty$, 
and $w=\varphi+it$, $\bar w=\varphi-it$ are complex coordinates on the
torus. The three-dimensional metric \footnote{$G$ represents Newton's
constant and $l$ is the anti-de Sitter radius related to a negative 
cosmological constant by $\Lambda = -1/l^2$. We set $\hbar=1$
throughout the paper.},  
\begin{equation}
 ds^2 = 4Gl( L dw^2 + \bar L d\bar w^2) +
\left(l^2 e^{2\rho} + 16 G^2 L \bar L e^{-2\rho} \right)dw d\bar w +
l^2 d\rho^2,
\label{dsLL}
\end{equation}
where $L=L(w)$ and $\bar L=\bar L(\bar w)$ are arbitrary functions of
their arguments satisfies the following properties: \\

\noindent {\bf (i) Exact solution} \\ 
The metric (\ref{dsLL}) is an exact solution to the three-dimensional
vacuum Einstein equations with a negative cosmological constant
$\Lambda=-1/l^2$. The leading and first subleading terms of
(\ref{dsLL}) (in
powers of $r=e^\rho$) are, of course, the ones dictated by the general
analysis of \cite{BH} for asymptotically anti-de Sitter spacetimes.
What is perhaps not so well-known is that adding the term $e^{-2\rho}
L\bar L$, the metric becomes an {\it exact} solution. Most
importantly, (\ref{dsLL}) is the most general solution, up to trivial
diffeomorphisms, which is asymptotically anti-de Sitter. Note that
since (\ref{dsLL}) contains two arbitrary functions, it gives rise to
an infinite number of solutions. 

We shall work in the Euclidean sector of the theory. This means that
$w$ is a complex coordinate related to the spacetime coordinates as
$w=\varphi+it$. The metric (\ref{dsLL}) is then complex. As we shall
see, this will not bring in any problems in the quantization. For real
values of $w$, the metric (\ref{dsLL}) is a solution to the
Minkowskian equations of motion.\\ 

\noindent {\bf (ii) Physical degrees of freedom}\\
Two solutions of the form (\ref{dsLL}) with different values for $L$
and $\bar L$ represent physically different configurations which
cannot be connected via a gauge transformation.  In the quantum
theory, where $L$ and $\bar L$ will become operators, different
expectations values for them will be associated to different
solutions.

This is a non-trivial statement.  Since (\ref{dsLL}) is a solution to
the three-dimensional Einstein equations it has constant curvature
and then, locally, is isometric to anti-de Sitter space
(see Eq. (\ref{adS}) below). The point here is that the coordinate
transformations which change the values of $L$ and $\bar L$ are not
generated by constraints and therefore they are not gauge symmetries.
This point will be analysed in detail in the Chern-Simons formulation
in section \ref{Global}, and in the metric formulation in Sec.
\ref{Diff/CS-GR}.  \\

\noindent {\bf (iii) Residual conformal symmetry}\\
The metric (\ref{dsLL}) has a residual conformal symmetry.  There
exists a change of coordinates $\{w,\bar w, \rho\} \rightarrow
\{w',\bar w', \rho'\}$ such that the new metric looks exactly like
(\ref{dsLL}) with new functions $L'$ and $\bar L'$. See Sec.
\ref{Diff/CS-GR} for the proof of this statement. This change of
coordinates is parametrized by two functions $\varepsilon(w)$ and 
$\bar\varepsilon(\bar w)$. The new function $L'$ is related to the old 
one via $L'=L + \delta L$ with,
\begin{eqnarray}
\delta L = i(\varepsilon \partial L + 2 \partial \varepsilon L -
{c\over 12} \partial^3 \varepsilon)
\label{deltaL}
\end{eqnarray}
where $c$ is given by
\begin{equation}
c= {3l \over 2G}.
\label{c}
\end{equation}
The same transformation holds for $\bar L$. Thus, under this symmetry,
$L$ and $\bar L$ are quasi-primary fields of conformal dimension 2. 
This symmetry, properly defined acting on the gravitational variables,
can be shown to be also a global symmetry of the action\cite{BH}. The
canonical generators are the functions $L$ and $\bar L$ themselves and
the associated algebra is the Virasoro algebra\cite{BH},
\begin{equation}
[L_n,L_m] = (n-m) L_{n+m} + {c \over 12} n^3 \delta_{n+m}, 
\label{Vir}
\end{equation}
where the central charge is defined in (\ref{c}) and,
\begin{equation}
L(w) = \sum_{n\in Z} L_n e^{inw}.
\label{Lmodes}
\end{equation}
We are using here a
non-standard form of the central term. The usual $n(n^2-1)$ form can
be obtained simply by shifting the $L_0$ mode as $L_0\rightarrow L_0 -
c/24$. This convention is appropriated to the black hole background
which has an exact $SO(2)\times SO(2)$ invariance. See
\cite{Coussaert-H} for a discussion on this point in the supergravity
context.\\ 

\noindent {\bf (iv) Asymptotic conformal symmetry}\\
The residual symmetry (\ref{deltaL}) is not an exact symmetry of any
background metric. 
Rather it is a symmetry of the space of solutions described by
(\ref{dsLL}); it maps one solution into another one. However, this
symmetry can be regarded as an {\it asymptotic} symmetry of anti-de
Sitter space because the $r\rightarrow\infty$ form of (\ref{dsLL}) is
asymptotic Euclidean adS$_3$ space (note the redefinition of
coordinates: $w=\varphi+it$, $r=le^\rho$),
\begin{equation}
ds^2_{(r\rightarrow\infty)} \rightarrow r^2 (dt^2 + d\varphi^2).
\label{asym/dsLL}
\end{equation}
(We have kept here only the leading terms in powers of $r$, but note
that there is also a term $2i(L-\bar L)dtd\varphi$ of order one which
is allowed by the boundary conditions\cite{BH}.) Since the asymptotic
behaviour (\ref{asym/dsLL}) does not see $L$ and $\bar L$ it is
invariant under the transformation (\ref{deltaL}). This symmetry was
discovered in \cite{BH}. See \cite{Strominger97,Navarro} for recent
discussions. \\

\noindent {\bf (v) Basic dynamical variables and induced Poisson
brackets}\\
Up to some global issues, the Virasoro algebra (\ref{Vir}) can be
regarded as the basic Poisson bracket algebra of the gauge-fixed
residual variables. In other words, the functions $L(w)$ and $\bar
L(\bar w)$ appearing in (\ref{dsLL}) are the part of the metric field
$g_{\mu\nu}(x^\mu)$ which survives after the gauge is fixed (i.e.,
after gauge conditions are imposed and the constraints solved), with
anti-de Sitter boundary conditions. The equal-time Poisson bracket of
general relativity, $\{\pi^{ij},g_{kl}\}=\delta^{ij}_{kl}$, induces
the Virasoro algebra (\ref{Vir}) on the residual dynamical functions
$L$ and $\bar L$. A technical note is convenient here. From the
dynamical point of view, $L$ and $\bar L$ both depend on $w$ and $\bar
w$. However, the gauge-fixed equations of motion read $\partial_{\bar
w} L=0$ and $\partial_{w} \bar L=0$ leading to $L=L(w)$ and $\bar
L=\bar L(\bar w)$. 

The identification of the Virasoro operators as basic variables is not
natural from the point of view of conformal field theory. In the
standard situation, the Virasoro algebra is associated to a symmetry
rather than to the basic commutator. (A good analogy is to consider
the angular momentum components $L_i$ as basic variables, satisfying
$[L_i,L_j] = i\epsilon_{ijk} L_k$, without knowing the existence of
$q^i,p_j$.)  One of the main problems of three-dimensional quantum
gravity is to identify what is the conformal field theory behind the
Brown-Henneaux conformal symmetry. Classically, Liouville theory
\cite{CHvD} seems to be a good candidate, however, its quantisation
does not give the right counting for the black hole entropy
degeneracy\cite{Carlip98-1}.  Treating the Virasoro algebra as basic
Poisson algebra is also well-motivated classically but it does not
give the right counting either (see Sec. \ref{Quantum-met}). We shall
discuss this issues in Sec. \ref{Quantum-met}, as well as two possible
modifications of the boundary dynamics which do provide the right
counting of states.   \\

\noindent {\bf (vi) Black holes and adS space}\\
The space of solutions described by (\ref{dsLL}) contains black holes. 
If $L$ and $\bar L$ are constants (no $w,\bar w$ dependence) with 
only $L_0$, $\bar L_0$ different from zero and parametrized as 
\begin{equation}
Ml=L_0+\bar L_0, \ \ \ \ \ \ \ J=L_0-\bar L_0,
\label{MJ}
\end{equation}
then the metric (\ref{dsLL}) is globally isometric to the Euclidean
three-dimensional black hole \cite{BTZ,BHTZ} of mass $M$ and angular
momentum $J$. Eq. (\ref{MJ}) means that the Virasoro operators vanish
on the vacuum black hole.  The corresponding algebra is (\ref{Vir}).
The Euclidean black hole metric in Schwarzschild coordinates reads 
\begin{equation}
ds^2 = l^2 N^2 dt^2+N^{-2}dr^2+r^2(d\varphi +i N^\varphi dt)^2, 
\label{BTZ}
\end{equation}
with 
\begin{eqnarray}
      N^2(r) &=& -8MG + {r^2 \over l^2} + {16G^2J^2 \over r^2}, \\
N^\varphi(r) &=& {4GJ \over r^2}.
\end{eqnarray}
See Sec. \ref{Sec-BTZ} for the explicit transition from (\ref{dsLL})
to (\ref{BTZ}).

For $M>|J|$, this metric has two horizons which are the solutions to
the equation $N^2(r_\pm)=0$.  It is often convenient to define
the Euclidean angular momentum $J_E$ as $J_E=iJ$ and then the $i$ in
(\ref{BTZ}) does not occur. Note also that in the Euclidean sector,
the black hole manifold does not see the interior $r<r_+$.  See Sec.
\ref{Sec-BTZ} for more details on the relation between (\ref{BTZ}) and
(\ref{dsLL}). Note that since the coordinates $w$ and $\bar w$ are
defined on a torus, the only globally well-defined solutions are the
ones with constant $L$ and $\bar L$. 

For $J=0$ and $8MG=-1$ the metric (\ref{BTZ}) reduces to
Euclidean anti-de Sitter space in three dimensions, 
\begin{equation}
ds^2_{adS} = l^2\left(1+ \frac{r^2}{l^2} \right)
dt^2 + \left(1+\frac{r^2}{l^2}\right)^{-1}dr^2 + r^2 d\varphi^2. 
\label{adS}
\end{equation} \\

\noindent {\bf (vii) A quantum metric, the `CFT$\rightarrow$ Geometry'
map and} \\ 
{\bf black hole entropy}\\
Once the functions $L$ and $\bar L$ are promoted to be operators
acting on Fock space, the metric (\ref{dsLL}) becomes a well-defined
operator, denoted as $d\hat{s}^2$, on that space.  We then find a map
from Fock's space (representations of the Virasoro algebra) into
the independent classical solutions of Einstein's equations. Let
$|\Psi>$ a state in Fock's space, we have 
\begin{equation}
|\Psi> \ \ \rightarrow \ \ ds^2_\Psi = <\Psi| d\hat s^2 | \Psi>
\label{map}
\end{equation}
with $L_\Psi =<\Psi| L |\Psi>$ and 
$\bar L_\psi =<\Psi| \bar L |\Psi>$. By construction
$ds^2_\Psi$ is a solution to the classical Einstein equations because
(\ref{dsLL}) is a solution for arbitrary functions $L$ and $\bar L$.
It also follows that the full set of states $|\Psi>$ generates the
full space of classical solutions.  Note that this map is valid
independently of the structure of the conformal field theory
generating the Virasoro algebra. We can then ask the question of how
many states are there in Fock space such that they induce through
(\ref{map}) a black hole of a mass $M$ and angular momentum $J$. 
The answer to this question of course depends on the structure of the
Hilbert space. We shall study this point in detail in Sec.
\ref{Quantum-met}.  \\   

\noindent {\bf (viii) Relation to 2d induced gravity}\\
Finally, note that for $\bar L=0$ and fixed $r$, the metric 
(\ref{dsLL}) is equal to Polyakov's \cite{Polyakov2d} 2d lightlike
metric which yields an $SL(2,\Re)$ algebra. Since 3d gravity is known
to induce 2d gravity at the boundary ($r$ fixed), the understanding of
the quantum properties of (\ref{dsLL}) may yield new information about
2d gravity. 

%-----------------------
\subsection{Organisation of the article} 

The goal of this article is to discuss and provide the proofs for the
above properties of the metric (\ref{dsLL}).  We have written
(i)-(viii) in a metric formulation of gravity because our final target
is quantum gravity. However, the explicit proofs will be given in
terms of the Chern-Simons formulation \cite{Achucarro-T} of
three-dimensional gravity because they are simpler and provide a rich
mathematical structure.  

In Sec. \ref{CS/GLOBAL} we give a short introduction to Chern-Simons
gravity and its phase space. A detailed discussion about boundary
degrees of freedom is included in that section.  In Sec.
\ref{KM/VIR} the explicit solution to the equations of motion, with
two different classes of boundary conditions, is written down (in
terms of the Chern-Simons fields) and their induced Poisson brackets
are displayed.  Finally, in Sec. \ref{Metric}, we go back to
the metric formulation and apply the results to quantum
three-dimensional gravity.

%----------------------------------------------------------
%----------------------------------------------------------
\section{Chern-Simons gravity and global degrees of freedom}
\label{CS/GLOBAL}

In this section we shall first briefly describe the Chern-Simons
formulation of three-dimensional gravity. Then we analyse the issue of
global degrees of freedom associated to the presence of boundaries. We
shall also show in this section (see Sec. \ref{Unitary}) how the
boundary conditions solve part of the unitary problems of
three-dimensional gravity.  

%------------------------
\subsection{Chern-Simons gravity, its equations of motion and their
solutions}

In our approach to the quantum black hole problem, the Chern-Simons
formulation of 2+1 gravity will be of great help. This formulation was
discovered in \cite{Achucarro-T} and its quantum properties (for
closed manifolds) were explored in \cite{Witten88}.  An extensive
treatment can be found in \cite{Carlip-book}. In a few words, the
Chern-Simons formulation is a field redefinition that
simplifies the equations and introduces a rich mathematical structure.  

The basic variables of general relativity in the tetrad formalism are
the triad $e^a$ and the spin connection\footnote{In three
dimensions one defines $\omega^a = (-1/2) \epsilon^a_{\ bc}
\omega^{bc}$. It follows that the 2-form curvature  $R^{ab} =
d\omega^{ab} + \omega^a_{\ c} \ww\omega^{cb}$ can be written in the
form $R^{ab} = -\epsilon^{ab}_{\ \ c} R^c$ with $R^a=d\omega^a + (1/2)
\epsilon^a_{\ bc} \omega^b \ww\omega^c$. In the same way, the
torsion $T^a=de^a+\omega^a_{\ b} \ww e^b$ reads $T^a = de^a +
\epsilon^a_{\ bc} \omega^b\ww e^c$. These definitions depend on the
signature. The formulae
displayed here are appropriated to Euclidean signature.} $\omega^a$.  
The equations of motion of three dimensional gravity with a negative
cosmological constant in these variables are simply
\begin{equation}
R^a = {1\over 2l^2} \epsilon^a_{\ bc}\, e^b \ww e^c, \ \ \ \ T^a =0.
\label{gr/eqn}
\end{equation}
We define now two new fields according to,  
\begin{equation}
A^a = \omega^a + {i \over l} e^a, \ \ \ \ \ \  
\bar A^a = \omega^a - {i\over l} e^a.  
\label{A/barA}
\end{equation}
The 1-form $A^a$ is an $SL(2,C)$ Yang-Mills gauge field. Let $F^a$ and
$\bar F^a$ the curvatures associated to $A^a$ and $\bar A^a$. The
discovery of Ach\'ucarro and Townsend \cite{Achucarro-T} is that the
equations,  
\begin{equation}
F^a =0, \ \ \ \ \ \bar F^a=0,  
\label{EqCS}
\end{equation}
are exactly equivalent to the three-dimensional Einstein equations
(\ref{gr/eqn}). Furthermore, the Einstein-Hilbert action is equal to
the combination,  
\begin{equation}
I[A,\bar A] = I[A]-I[\bar A],
\label{IAA}
\end{equation}
where $I[A]$ is the Chern-Simons action,
\begin{equation}
I[A] = \frac{k}{4\pi} \int \mbox{Tr} (A dA + \frac{2}{3} A^3).
\label{ICS}
\end{equation}
 
To determine the Chern-Simons coupling constant, or level, $k$ as a
function of the gravitational constants $G$ and $l$ we need to fix 
the representation of $A$ and $\bar A$. We use the anti-Hermitian
$SU(2)$ generators,  
\begin{equation}
J_1 = {i\over 2} \left( \begin{array}{cc}   0 &  1  \\
                            1 &  0   \end{array} \right), \ \ \
J_2 = {1\over 2} \left( \begin{array}{cc}   0 &  -1  \\
                           1 &  0   \end{array} \right), \ \ \
J_3 = {i\over 2} \left( \begin{array}{cc}   1 &  0  \\
                            0 &  -1   \end{array} \right),
\label{conventions}							
\end{equation}
which satisfy $[J_a,J_b]=\epsilon_{ab}^{\ \ c} J_c$ and Tr$(J_a J_b) 
= -(1/2) \delta_{ab}$ and define
\begin{equation}
 A=A^aJ_a, \quad  \bar A = \bar A^a J_a.
\label{AbarA}
\end{equation}
Note that we use the same $J$'s in both cases. This means that $\bar
A$ is not the complex conjugate of $A$. With these conventions,
comparing the Chern-Simons action with the Einstein-Hilbert action one
finds 
\begin{equation}
k =- {l \over 4G}.
\label{k}
\end{equation}
The sign of $k$ depends on the identity $\sqrt{g}=\pm e$ where $e$ is
the determinant of the triad. This sign determines the relative
orientation of the coordinate and orthonormal basis. We have chosen
here the plus sign which means that we work with $e>0$. 

Note that the gauge field $A$ is complex and thus the relevant group
is $SL(2,C)$ which is non-compact. This means that we cannot apply in
a straightforward way the quantization of Chern-Simons theory
described in \cite{Witten89}. Our prescription to define the quantum
theory will be to first find the general solution the classical
equations of motion, under prescribed boundary conditions, and then
quantize that space. As we
shall see, the boundary conditions will play a key role in making the
quantum theory unitary (see Sec. \ref{Unitary}).  

The convenience of the Chern-Simons formulation is evident. Instead of
working with a second order action in terms of the metric, we work
with two flat Yang-Mills fields. The equations (\ref{EqCS}) show
clearly that 2+1 gravity does not have any local degrees of freedom.
This means that all dynamics is contained in the holonomies
\cite{Witten88} and boundary degrees of freedom
\cite{Carlip95,Balachandran2}.    

The general solution to the equations (\ref{EqCS}) can be written in
the form,  
\begin{equation}
A = g^{-1} dg + g^{-1} H g,
\label{sol/g}
\end{equation} 
where $H$ is also flat ($dH + H\w H=0$) but cannot be written as
$u^{-1} du$ with $u$ single valued.  Similar arguments hold for $\bar
A$. The group element $g(x)$ is a single valued map from the manifold
to the group.  The space of solutions (\ref{sol/g}) is
invariant under 
\begin{equation}
A \rightarrow A' = U^{-1} A U + U^{-1} dU
\label{gauge/U}
\end{equation}
where $U$ is another map from the manifold to the group. In principle
(see below for a detailed discussion), we can use this symmetry to set
$g=1$ in (\ref{sol/g}) and thus all solutions are classified only by
the independent values of holonomy $H$.  The quantization of
this sector of phase space was first discussed in \cite{Witten88}. Its 
dimensionality is finite and cannot account for the large black hole
degeneracy.  For this reason we shall not
consider them here anymore. However, it is important to stress that
the black hole gauge field does have non-trivial holonomies. Indeed,
it can be shown that the gauge field corresponding to a black hole
satisfies \cite{Cangemi-LM},
\begin{equation}
\mbox{P} \exp \oint A =:\exp(w), \ \ \ \ \ \mbox{P} \exp \oint
\bar A =:\exp(\bar w)
\end{equation}
where
\begin{equation}
\mbox{Tr}( w^2 + \bar w^2) = 32\pi^2 MG,  \ \ \ \ \ 
\mbox{Tr}( w^2 - \bar w^2) = {32\pi^2 JG\over l},
\end{equation}
and $M$ and $J$ are the black hole mass and angular momentum, 
respectively. Only for $8GM=-1$ and $J=0$ these holonomies are
trivial. The corresponding solution is anti-de Sitter space
(\ref{adS}).

%-----------------------------------
\subsection{The strategy}

Since Chern-Simons theory does not have any local excitations, all
relevant degrees of freedom are global.  We shall consider here the
situation on which the topology is fixed and thus all relevant states
come from the presence of boundaries. The existence of boundary
degrees of freedom has been analysed in great detail by Carlip
\cite{Carlip95-96} using a covariant formalism and path integrals (see
also \cite{Balachandran2} for an approach similar to ours). Here, we
shall describe an equivalent procedure based on the Hamiltonian
formalism in the form discussed by Regge and Teitelboim
\cite{Regge-CT}. This approach can be summarised in the following
steps. Given an action $I[\phi]$ with a gauge symmetry $\delta_G \phi$
we need to: 
\begin{itemize}
\item
Impose boundary conditions on the fields such that $\delta 
I[\phi]/\delta \phi$ exists. These boundary conditions are not 
unique and their election represent an important physical input
into the theory.  [In practice, one first decides the boundary
conditions and then add to the action the necessary boundary terms to
make it differentiable.] 
\item
Find the sub-group of gauge transformations that leave the boundary
conditions and the action invariant. 
\item
Find the canonical generators which generate the symmetries of the
action. If a generator is a constraint, we shall call the associated
symmetry a {\it gauge} symmetry. Configurations which differ by gauge
symmetries are identified and represent the same physical state. 
Conversely, if a generator is different from zero (even on-shell), we
call the associated symmetry a {\it global} symmetry. Note that
according to this definition, global symmetries do not need to be
rigid. Global symmetries map the space of physical states into itself.   
\end{itemize}

We shall see that a proper distinction between global and gauge
symmetries is crucial to understand the boundary degrees of freedom in
Chern-Simons theories.

%-------------------------------------
\subsection{Global symmetries and boundary degrees of freedom}   
\label{Global}

The appearance of boundary degrees of freedom can be summarized as
follows. Chern-Simons theory has $3N$ fields $A^a_\mu$
($\mu=0,1,2$;~ $a=1,...,N$). However, the gauge symmetries of the
action tells us that they do not represent independent physical
degrees of freedom. In fact, locally, using the symmetry
(\ref{gauge/U}) one can kill all of them (the temporal component
$A^a_0$ is a Lagrange multiplier, while the spatial components $A^a_i$
are $2N$ fields subject to $N$ constraints $F^a_{ij}=0$ plus $N$ gauge
conditions). 

The question we want to address here is whether the symmetry
(\ref{gauge/U}) is really a gauge symmetry, in the sense that two
fields related by it are to be considered the same, or not. After
properly defining what a gauge transformation is, we shall see that at
the boundary the transformation (\ref{gauge/U}) is not a gauge
symmetry, although it is still a symmetry of the action.  Then, two
solutions of the form (\ref{sol/g}) with $g$ and $g'$ such that, at
the boundary, $g \neq g'$ represent two different physical
configurations. This boundary effect can give rise to an infinite
number of degrees of freedom (independent solutions to the equations
of motion).  At this point we can make contact with Carlip's
would-be-gauge degrees of freedom approach: the field $g$, at the
boundary, is dynamical and its dynamics is governed by a $WZW$
action\cite{Carlip95-96}.    

In the presence of boundaries the definition of a gauge symmetry
becomes delicate because not all the transformations encoded in
(\ref{gauge/U}) are generated by constraints. Indeed, if $U$ does not
approach the identity map at the boundary, then the associated
canonical generator is a non-zero quantity and hence that
transformation is not a gauge symmetry. See Sec. \ref{affine} for 
a proof of this statement in Chern-Simons theory.  

Following Dirac's quantization procedure (see \cite{HT-book} for an
extensive treatment), we define a {\it gauge} transformation as a
symmetry generated by a (first class) constraint. On the contrary, a
symmetry of the action generated by a non-zero quantity is called
{\it global}, even if it is not rigid. By definition, the space of
physical states, or phase space, is the set of fields which satisfy
the equations of motion, modulo gauge transformations. Let us ignore
the holonomies for a moment. The general solution (\ref{sol/g}) then 
reduces to $A=g^{-1}dg$. If no boundaries are present, this space of
solutions is trivial containing only one element $A=0$ because the
transformations (\ref{gauge/U}) are generated by constraints (hence, 
they represent gauge symmetries) and one can use (\ref{gauge/U}) to
set $g=1$ and thus $A=0$.    

On the contrary, if there is a boundary, part of the symmetry
(\ref{gauge/U}) is not generated by a constraint (to be proved below).
Therefore, while it is still true that we can transform any flat
$A$ to 0 using (\ref{gauge/U}), it is not true that the state $A$ and
the state $0$ represent the same physical configuration.  Both states, 
$A\neq 0$ and $A=0$ (at the boundary), are solutions to the equations
of motion and they are related by a symmetry of the action. However,
they are physically distinguishable. Indeed, there is a gauge
invariant conserved charge which takes different values in each state.
Our main problem is then to determine the set of fields $\hat A$ 
which solve the equations of motion and cannot be set to zero by the
action of a constraint. As we shall see, in Chern-Simons theory there
is an infinite number of them. 

In a quantum mechanical notation, the above discussion can be
summarized as follows. Denote by $G_0$ the set of transformations
which are true gauge symmetries generated by constraints, and by $Q$
those which are not. Physical states satisfy $G_0|\Psi>=0$. On the
other hand, $Q$ generates a symmetry of the space of
physical states, that is $Q|\Psi>=|\Psi'> $.  We shall prove
explicitly (at least in Chern-Simons theory; for a general
discussion see \cite{Brown-Henneaux1}) that $G_0$ and $Q$ satisfy an
algebra of the form,
\begin{eqnarray}
~[G_0,G_0] &=& G_0, \label{00} \\ 
~[G_0,Q] &=& G_0, \label{0G} \\
~[Q,Q] &=& Q + c \label{GG}
\end{eqnarray}
where $c$ represents (schematically) a possible central term. Eq.
(\ref{00}) is the definition of first class constraints. Eq.
(\ref{0G}) means that if $|\Psi>$ is physical ($G_0|\Psi>=0$) then
$Q|\Psi>$ is also physical ($Q$ generates a global symmetry of
the Hilbert space).  Finally, Eq. (\ref{GG}) is the algebra of the
globally symmetry. The appearance of central terms in (\ref{GG})
cannot be discarded by a general principle \cite{Brown-Henneaux1}.
Note however that since $Q$ does not generate a gauge symmetry and it
is different from zero, the central term does not represent any
trouble after quantization. An interesting and important example on
which the central term is present was discovered in \cite{BH}. 

The above discussion is a quick summary of the results presented in
\cite{Regge-CT,Benguria-CT,Brown-Henneaux1,BH}, and many other papers
that have followed this work. The nice property of Chern-Simons theory
is that these ideas can be tested with minimum calculations. Another
system which is simple to analyse is Yang-Mills theory on which the
above analysis leads to the definition of global colour
charges\cite{Abbot-D}. However, in that case, the resulting global
algebra is finite dimensional and does not have any central terms.

%---------------------------------------
\subsection{Boundary conditions in Chern-Simons gravity}
\subsubsection{Making the action differentiable}

The black hole manifold is asymptotically anti-de Sitter and then it
has a boundary. In the Euclidean sector, the boundary has the topology
of a torus with compact coordinates $\varphi$ and $t$. It is
convenient to define the complex coordinates on the torus
\begin{equation}
w = \varphi + it, \ \ \ \ \ \bar w = \varphi - it,  
\label{w}
\end{equation}
and then $A_\varphi d\varphi + A_t dt = A_w dw + A_{\bar w} d\bar w$. 

Boundary conditions are necessary in order to ensure that the action
principle has well defined variations. As discussed above, all the
dynamics of 2+1 gravity is contained in the boundary conditions. For
this reason, it is a key problem to choose them judiciously.  In
particular, if they are too strong there will be no dynamics left in
the theory.  For the black hole problem (which is asymptotically
anti-de Sitter) there is a natural choice of boundary conditions first
discussed in \cite{CHvD} in the Minkowskian signature and extended to
Euclidean signature in \cite{BBO}. In the coordinates (\ref{w}) they
read simply  
\begin{equation}
A^a_{\bar w}=0, \ \ \ \ \bar A^a_w=0  \ \ \ \ \ \mbox{(at the
boundary)}.
\label{bcond}
\end{equation}

A quick way to convince ourselves that the black hole  satisfies this
condition is to consider the constant curvature metric 
\begin{equation}
ds^2 = e^{2\rho} (dx^2 + dy^2) + l^2d\rho^2.
\label{cc}
\end{equation}
A natural election for the triads is $\{e^1= e^\rho dx,e^2=e^\rho dy,
e^3=ld\rho\}$. The torsion equation $de^a+ \epsilon^a_{\ bc} \omega^b
\wedge e^c =0$ yields for the components of $\omega^a$ $\{\omega^1=
-(1/l) e^\rho dy$, $\omega^2 = (1/l)e^\rho dx, \omega^3 = 0\}$.
Defining $w=x+iy$ it is clear that $A^a = \omega^a + (i/l) e^a$ and
$\bar A^a = \omega^a - (i/l) e^a$ satisfy (\ref{bcond}).  It can be
shown that the black hole metric (\ref{BTZ}) which is also of constant
curvature satisfies (\ref{bcond}) as well \cite{BBO}.  See
\cite{Carlip-T} for the explicit transition from (\ref{cc}) to
(\ref{BTZ}). 

This example also illustrates the choice of orientation. Suppose we
choose a new set of triads given by $\tilde e^1= -e^\rho dx$, $\tilde
e^2=-e^\rho dy$ and $\tilde e^3 = -ld\rho$. These new fields  satisfy
the torsion equation because is homogeneos in $e$, and Einstein
equations because they are quadratic in $e$. However, the determinant
of $\tilde e^a_\mu$ is negative. As we remarked before, the value of
$k$ given in (\ref{k}) depends on the orientation of the orthonormal
basis, and the identity $\sqrt{g}=\pm e$. We have chosen $e>0$ and
then the election $\tilde e^a$ is not allowed.   

Let us check that (\ref{bcond}) are enough to make the action
differentiable. The variation of the Chern-Simons action gives a term
proportional to the equations of motion plus a boundary term,
\begin{eqnarray}
\delta I_{CS} &=& \int_M (\mbox{eom})_a \delta A^a  +  
   \frac{k}{4\pi} \int_{\partial M} g_{ab}A^a \wedge \delta A^b
   \nonumber\\ 
              &=&\int_M (\mbox{eom})_a \delta A^a  +
              \frac{k}{4\pi} \int_{\partial M} g_{ab} (A^a_w \delta 
			     A^b_{\bar w} - A^a_{\bar w} \delta A^b_{w})
			     \label{dI}\\
              &=&  \int_M (\mbox{eom})_a \delta A^a  +  0.
              \nonumber
\end{eqnarray}
The boundary term vanishes due to (\ref{bcond}).  Thus, the variation
of the action under the boundary condition (\ref{bcond}) is well
defined. Later we will restrict further the values of the gauge field
at then boundary, but for the purposes of this discussion the above
boundary conditions are very useful. 

In summary, we work with the Chern-Simons action (\ref{IAA})
supplemented with the boundary conditions (\ref{bcond}) and no added
boundary terms. [Note that when passing to the Hamiltonian formalism
there will be a boundary term\cite{BBO}.]  As a further check that
this action is appropriated to the black hole problem,
one can prove \cite{BM} that its value on the Euclidean black hole
solution is finite and gives the right canonical free energy
(Gibbons-Hawking approximation).      

%----------------------------------------
\subsubsection{The chiral boundary group}
\label{Boundary-group}

The second step in the Regge-Teitelboim procedure is to determined how
the gauge symmetries are affected by the boundary conditions, i.e., to
determine the residual group of transformations that preserves
(\ref{bcond}).  This is actually very simple. We look for the set of
parameters $\lambda^a$ satisfying 
\begin{equation}
\delta A^a_{\bar w} = D_{\bar w} \lambda^a =0  \ \ \ \ \ \mbox{(at the
boundary)}.
\label{bcond/p}
\end{equation}
Since by (\ref{bcond}) $A_{\bar w}=0$ this condition simply imply
that $\partial_{\bar w} \lambda^a=0$. The subset of gauge
transformations  leaving (\ref{bcond}) invariant are then those whose
parameters at the boundary are chiral, only depend on $w$. 

Let us now check that this group leaves the action invariant. The
variation of the Chern-Simons action under $\delta A^a = D\lambda^a$
gives a boundary term,
\begin{eqnarray}
\delta I_{CS} &=&\frac{k}{4\pi} \int_{\partial M} A^a \wedge
D\lambda_a  \nonumber\\ 
              &=&\frac{k}{4\pi} \int_{\partial M} (A^a_w D_{\bar
              w}\lambda_a - A^a_{\bar w}D_w \lambda_a) \label{dI2}\\
              &=& 0 \nonumber
\end{eqnarray}
which vanishes thanks to (\ref{bcond}) and (\ref{bcond/p}). 
There is an important point to be stressed here. It is often said in
the literature that the Chern-Simons action is invariant under $\delta
A^a = D\lambda^a$ only if $\lambda=0$ at the boundary. This is, as we
have just shown, not true.  The right statement is that $\lambda$
cannot be completely arbitrary at the boundary but it can be different
from zero. Under the boundary condition (\ref{bcond}),
the action is invariant under transformations with non-zero values of
$\lambda^a$ at the boundary provided that parameter is chiral
($\lambda^a=\lambda^a(w)$). This gives rise to an infinite dimensional
symmetry.

%-----------
\subsection{Affine (Kac-Moody) algebras}
\label{affine}

Let us briefly describe the main steps leading to (\ref{00}-\ref{GG})
in Chern-Simons theory. For more details, the reader is referred to
\cite{Balachandran1,B}.   

In the 2+1 decomposition of the gauge field $A^a = A^a_0 dt + A^a_i
dx^i$, the Chern-Simons action reads,
\begin{equation}
I[A_i,A_0] = \frac{k}{8\pi} \int dt \int_\Sigma  \epsilon^{ij}
\delta_{ab} 
(A^a_i \dot A_i^b - A^a_0 F^b_{ij} )  + B,
\end{equation}
where $B$ is a boundary term. Here we have used that
Tr$(J_aJ_b)=-(1/2) \delta_{ab}$. The coordinates $x^i$ are local
coordinates on the spatial surface denoted by $\Sigma$. This
action has $2N$ dynamical fields $A^a_i$ ($a=1,...,N$; $i=1,2$) and
$N$ Lagrange multipliers $A^a_0$. The dynamical fields satisfy the
basic equal-time Poisson bracket
algebra,
\begin{equation}
\{A^a_i(x),A^b_j(y) \} = \frac{4\pi}{k} \epsilon_{ij} \delta^{ab}
\delta^2(x,y).
\label{pAA}
\end{equation}
The Poisson bracket of two functions $F(A_i)$ and $H(A_i)$ is 
computed as
\begin{equation}
\{F,H\} = \frac{4\pi}{k} \int_\Sigma d^2 z 
{\delta F \over \delta A^a_i(z)} \epsilon_{ij} \delta^{ab} {\delta
H \over \delta A^b_j(z)}.
\label{pFH}
\end{equation}
The functionals $F$ and $H$ need to be differentiable with respect to
$A_i$.

The equation of motion with respect to $A_0$ leads to the constraint
equation,
\begin{equation}
G_0^a = {k \over 8\pi} \epsilon^{ij} F^a_{ij} \approx 0,  
\label{G=0}
\end{equation}
which, we expect, will be the canonical generator of the gauge
transformations $\delta A^a_i = D_i \lambda^a$. This is indeed true
but only for those transformation whose parameters vanish at the
boundary. Indeed, define $G_0(\lambda) = \int_\Sigma  \lambda_a G_0^a$
and compute $\delta_\lambda A_i(x) = [A^a_i(x), G_0(\lambda)]$. It is
direct to see that the functional derivative of $G_0(\lambda)$ with
respect to $A_i$ is well-defined only if $\lambda^a$ vanishes at the
boundary. In that case, one does find $[A^a_i(x), G_0(\lambda)] = D_i
\lambda^a$ and thus $G_0(\lambda)$ generates the correct gauge
transformation. (We stress here that $G_0(\lambda)$  should not be
identified with the constraint: $G_0(\lambda)$ is the constraint
smeared with a parameter that vanishes at the boundary.)

However, as we discussed in Sec. \ref{Boundary-group},  the
Chern-Simons action with the boundary condition (\ref{bcond}) is also
invariant under transformations whose parameters at the boundary are
chiral $\lambda^a=\lambda^a(w)$ but different from zero. What is then
the generator of those transformations? Consider 
\begin{equation}
Q(\lambda) = \int_{\Sigma} \lambda_a G_0^a - {k \over 4\pi}
\int_{\partial \Sigma}
\lambda_a A^a.
\label{Q}
\end{equation}
It is easy to check that the boundary term arising when varying the
bulk part of (\ref{Q}) is cancelled by the boundary term, without
imposing any conditions over $\lambda$. The combination (\ref{Q}) then
has well defined variations even if $\lambda$ does not vanish at the
boundary.  Furthermore, one can check that $[A^a_i(x), Q(\lambda)] =
D_i \lambda^a$ and therefore $Q$ indeed generates those
transformations whose parameters do not vanish at the boundary. 

The key point here is that $Q$ is no longer a combination of the
constraints (for $\lambda^a|_{\partial\Sigma} \neq 0$) and thus it is
different from zero, even on-shell. According to the previous
discussion, $Q$ generates a global symmetry of the action. Two
configurations which differ by a transformation generated by it
represent physically different states. As a direct application of this
result, we find that two flat connections $A$ and $A'$ whose values at
the boundary differ, $(A-A')|_{\partial \Sigma} \neq 0$, cannot be
connected by the action of a constraint. Thus, as we have anticipated,
the values of $A$ at the boundary represent the physically relevant
degrees of freedom. The next step is to prove that there exists
solutions to the equations of motion, satisfying the boundary
conditions (\ref{bcond}), with different values for $A$ at the
boundary. This is done in the next section.   

By direct application of the Poisson bracket (\ref{pFH}) one can find
the algebra of two transformations with parameters $\eta$ and
$\lambda$ not vanishing at the boundary,
\begin{equation}
[Q(\eta),Q(\lambda) ] = Q([\eta,\lambda]) + \frac{k}{4\pi}
\int_{\partial \Sigma} \eta_a d\lambda^a 
\label{KM'}
\end{equation}
where $[\eta,\lambda]^a=\epsilon^a_{\ bc} \eta^b\lambda^c$. This
equation should be compared with (\ref{GG}). Also, note that if
$\lambda$ vanishes at the boundary then $Q(\lambda) = G_0(\lambda)$.
One can then easily see that (\ref{KM'}) reproduces (\ref{00}) and
(\ref{0G}) as well.    

The algebra (\ref{KM'}) provides the simplest way to determine the
Poisson bracket structure on the space of functions which cannot be
set to zero by the action of a constraint.  We shall do this
explicitly in next section.

%--------------------------
\subsection{Unitarity. An $SU(2)$ field}
\label{Unitary}

To end this section, we mention an important consequence of the
boundary conditions (\ref{bcond}). Namely, they provide a simple
solution to one of the problems with unitarity in Chern-Simons
gravity. As we have 
mentioned above, the gauge field $A=\omega^a + (i/l) e^a$ is complex
and therefore the relevant group is $SL(2,C)$ which is non-compact. It
has been argued in \cite{Witten91}, and explicitly used for example in
\cite{Carlip97} and \cite{BBO}, that under some conditions in the path
integral one can set $e^a=0$ and work with the $SU(2)$ gauge field
$A^a = \omega^a$. For closed manifolds, this has been shown to give a
good prescription \cite{Witten91}, but it is not the case for
manifolds with a boundary.    

The boundary conditions (\ref{bcond}) lead to a simple solution to
part of this problem. Indeed, expressing (\ref{bcond}) explicitly in
terms the triad and spin connection one finds,
\begin{equation}
\omega^a_{\bar w} + {i\over l} e_{\bar w} = 0, \ \ \ \ \ \ \ \  
\omega^a_w - {i\over l} e_w = 0.
\label{bcond/w}
\end{equation}
Using these equations, the non-zero components of $A_\mu$ and $\bar
A_\mu$ at the boundary, namely $A_w$ and $\bar A_{\bar w}$, can be
written in terms of the spin connection as,
\begin{equation}
A^a_w=2\omega^a_w, \ \ \ \ \ \ \bar A^a_{\bar w}=2\omega^a_{\bar w}.
\label{Aw}
\end{equation}
This shows that the non-zero components of the gauge field at the
boundary, which in fact carry all the dynamics, are real $SU(2)$
currents.   

Since all the dynamics of Chern-Simons theory will be
defined at the boundary, this simple observation means that we can
indeed work with two $SU(2)$ currents and forget about the non-compact
nature of $SL(2,C)$. Of course  in the bulk we still have an
$SL(2,C)$ field and this makes the statement ``all the dynamics is
contained at the boundary" delicate.  We shall not discussed this
point anymore in this paper. Our prescription will be to treat the
gauge degrees of freedom classically and to quantise the reduced phase
space, after the gauge has been fixed. In the language of
\cite{Moore-S}, the Chiral $WZW$ action arises classically (by varying
with respect to $A_0$ instead of integrating over it) and we work
with its basic Poisson bracket which is the $SU(2)$ affine algebra.
Actually, we shall not make explicit use of \cite{Moore-S}, but
rederive the same algebras by studying global symmetries of the
Chern-Simons action.  Some comments on the relation between both
methods will be given in Sec. \ref{WZW}.      

For later convenience, we mention here that using (\ref{bcond/w}) the
formulae (\ref{Aw}) can also be rewritten in terms of the triad as
\begin{equation}
A^a_w = {2 i \over l} e^a_w, \ \ \ \ \ \ \ \ 
\bar A^a_{\bar w}=-{2 i \over l} e^a_{\bar w}. 
\label{Ae}
\end{equation}
These formulae are more useful when constructing the metric out of the
connection via $g_{\mu\nu} = e^a_\mu e^b_\nu \delta_{ab}$. 

Note finally that the equations (\ref{bcond/w}) make a link between
the $i$ appearing in $A= w +(i/l) e$ and the $i$ appearing in the
complex structure given to the torus through $w=\varphi+\tau t$ (we
are using here $\tau=i$).     

%-----------------------------
%-----------------------------------------------
\section{The affine and anti-de Sitter solutions}  
\label{KM/VIR}

As discussed in detail in the last section, the presence of a 
boundary means that not all the values of the field $A$ are related by
proper gauge transformations.  Two solutions of the equations of
motion $A$ and $A'$ whose values at the boundary differ by a chiral
non-zero transformation are physically distinguishable solutions.  

This means that the space of solutions is not trivial.  In this
section we shall explicitly solve the equations of motion with the
boundary conditions (\ref{bcond}) and isolate the variables which are
physically relevant. We shall also find the induced Poisson bracket
acting on the space of dynamical (gauge-fixed residual functions)
degrees of freedom.   

%---------------------
\subsection{The affine solution} 
\label{Sec-KM}

We work on the solid torus with coordinates $\{w,\bar w, \rho\}$ and
$A=A_w dw + A_{\bar w} d\bar w + A_\rho d\rho$.  Our goal in this
section is to find the general solution to the equations of motion
which satisfies the boundary condition (\ref{bcond}).

The first step is to fix the gauge and eliminate the redundant degrees
of freedom. We impose the gauge condition,
\begin{equation}
A_\rho = i J_3,
\label{Arho}
\end{equation}
which implies that $\rho$ is a proper radial coordinate (see below).
Our conventions for the matrices $J_a$ are given in
(\ref{conventions}).  Note that the $i$ present in (\ref{Arho}) means
that $e_\rho \neq 0$. This is necessary because otherwise the triad
would be degenerate.  Next we impose the boundary condition $A_{\bar
w}=0$ [see (\ref{bcond})]. The equations of motion $F=0$ in
the coordinates  $\{w,\bar w, \rho\}$ read explicitly,
\begin{eqnarray}
\partial_\rho A_w - \partial_w A_\rho + [A_\rho,A_w]&=&0,\nonumber \\
\partial_\rho A_{\bar w} - \partial_{\bar w} A_\rho + [A_\rho,A_{\bar
              w}] &=& 0,     \nonumber\\
\partial_w  A_{\bar w}-\partial_{\bar w} A_w+[A_w,A_{\bar w}]&=&  0. 
 \label{eqns}      
\end{eqnarray}
The general solution to these equations in the gauge (\ref{Arho}) 
and satisfying $A_{\bar w}=0$ at $\partial M$ is,  
\begin{eqnarray}
A_w &=& b^{-1} \hat A(w) b, \nonumber\\
A_{\bar w} &=& 0, \nonumber\\
A_\rho &=& b^{-1} \partial_\rho b
\label{sol/KM}
\end{eqnarray}
where $\hat A$ is a chiral function, $\hat A=\hat A(w)$, but otherwise
arbitrary and  
\begin{equation}
b = e^{i\rho J_3} = \left( 
\begin{array}{cc}   
          e^{-\rho/2} &  0  \\
                    0 &   e^{\rho/2}  
\end{array} \right).
\label{b}
\end{equation}

Since $\hat A(w)$ is arbitrary, the space of solutions
(\ref{sol/KM}) is infinite dimensional. The question is whether
different solutions with different values for $\hat A$ are related by
gauge transformations or not. Consider a configuration of the form
(\ref{sol/KM}) and act on it with the transformation,
\begin{equation}
\delta A_\mu = D_\mu \eta,  \ \ \ \ \ \ \  \eta=b^{-1} \hat \eta(w) b.
\label{res/KM}
\end{equation}
It is direct to see that the effect of this transformation on the
solution is to produce another solution of the form (\ref{sol/KM})
with $\hat A' = \hat A + \hat D_w \hat \eta$ (here $\hat D_w \hat
\eta$ denotes covariant derivative in $\hat A$,  $\hat D_w \hat \eta =
\partial_w \hat\eta + [\hat A, \hat\eta]$). Thus, by acting on
(\ref{sol/KM}) with (\ref{res/KM}) we move around on the space of
solutions.  Actually, the transformations (\ref{res/KM}) are the most
general set of transformations that leave the boundary condition and
gauge fixing conditions invariant.

Since the parameter $\hat\eta$ appearing in (\ref{res/KM}) does not
vanish  at the boundary, the canonical generator of (\ref{res/KM}) is
a non-zero quantity of the form (\ref{Q}).  We then conclude that
different values of the function $\hat A$ are connected by global
transformations generated by (\ref{Q}) and not by the action of
constraints. The function $\hat A$ then represents dynamical degrees
of freedom. Even more, since (\ref{sol/KM}) is the most general
solution to the equations of motion with the boundary condition
(\ref{bcond}), the function $\hat A$ generates the full space of
non-trivial solutions with those boundary conditions.   

As it was pointed out in \cite{BBO}, this analysis is also valid if
the boundary is located at a finite value of the radial coordinate.
This observation is important, for example, if one expects to find a
conformal field theory at the black hole horizon. The main difference
between the metric and Chern-Simons approaches to global symmetries is
that in the former one works with diffeomorphisms while in the latter
with gauge transformations. Contrary to gauge transformations,
diffeomorphisms are not local and move the position of the boundary.
For example, the residual diffeomorphisms associated to anti-de
Sitter space \cite{BH} have a non-zero radial component. 

Our next step is to determine the Poisson bracket structure acting on
the space of solutions, that is, the induced Poisson bracket acting on
the functions $\hat A$. This is very easy thanks to a general
theorem proved in \cite{Brown-Henneaux1}.  First, we note that after
the gauge is fixed (the constraint is solved and (\ref{Arho}) is
imposed) the value of $Q$ given in (\ref{Q}) reduces to the boundary
term, 
\begin{equation}
\hat Q(\hat\eta) = -{k\over 4\pi} \int \hat \eta_a \hat A^a,
\label{Q'}
\end{equation}
which is, as we have emphasized, different from zero. The theorem
\cite{Brown-Henneaux1} states that after the gauge is fixed and one
works with the induced Poisson bracket (or Dirac bracket), the charge
$\hat Q$ satisfies the same algebra (\ref{KM'}) as it did the full
charge $Q$, 
\begin{equation}
[\hat Q(\hat \eta), \hat Q(\hat \lambda)]^*= \hat Q([\hat
\eta,\hat\lambda]) + {k  \over 4\pi } \int_{\partial \Sigma}
\hat\eta_a d \hat\lambda^a
\label{hatQQ}
\end{equation}
This algebra can be put in a more explicit form by defining, 
\begin{equation}
\hat A^a(w) = {2\over k} \sum_{n\in Z} T^a_n e^{inw}.
\label{T}
\end{equation}
One finds  
\begin{equation}
[T^a_n,T^b_m]^* = -\epsilon^{ab}_{\ \ c} T^c_{n+m} + {ink\over 2}
\delta^{ab}\delta_{n+m,0} .
\label{KM/PB}
\end{equation}
The algebra (\ref{KM/PB}) is called Kac-Moody or affine algebra and
represents an infinite dimensional symmetry of the space of solutions.
The quantum version is obtained simply by replacing the Dirac bracket
by $-i$ times the commutator,
\begin{equation}
[T^a_n,T^b_m] = i\epsilon^{ab}_{\ \ c} T^c_{n+m} + {nk\over 2}
\delta^{ab}\delta_{n+m,0}. 
\label{KM}
\end{equation}
This equation represents the algebra of the gauge-fixed basic
variables (analogous to $[q,p]=i$) of Chern-Simons theory with the
boundary condition (\ref{bcond}). There are various ways to see this
explicitly, which also provide alternative derivations of (\ref{KM}).
Conceptually, the most direct derivation of this result is by starting
with the three-dimensional Poisson bracket (\ref{pAA}). Then we fix
the gauge as in (\ref{Arho}) and solve the constraint
$F_{\rho\varphi}=0$.  The solutions to the constraint equation in this
gauge are parametrized by the function $\hat A$. One can then compute
the Dirac bracket of $\hat A$ with itself and find the affine algebra
(\ref{KM}).  Other methods yielding the same result are the $WZW$
approach followed in \cite{Moore-S}, and the symplectic method
\cite{Park}. The idea of looking at first class quantities and their
algebra in Chern-Simons theory was first discussed in
\cite{Balachandran1}. The presence of central terms in the algebra of
global charges in Chern-Simons theory was first discussed in
\cite{Mickelsson}. 

Note that in our applications to general relativity, the value of $k$
is negative (see Eq. (\ref{k})). This can be cured simply by 
replacing\footnote{We thank M. Asorey and F. Falceto for useful
conversation on this point.} $T^a_n \rightarrow T^a_{-n}$, or in other
words, by replacing the maximum weight condition $T^a_{-n}|0>=0$   by
$T^a_n|0>=0$ ($n>0$).  In any case, we shall not consider
(\ref{KM}) as the starting point for quantization but rather a
restriction of it which yields the Virasoro algebra with central
charge $c=-6k$. For this reason, in this particular calculation, it
seems convenient to work with a negative $k$. Incidentally, note that
the Virasoro algebra is also invariant under $L_n\rightarrow -L_{-n}$,
$c\rightarrow -c$.

Because of its affine symmetry we shall call the solution
(\ref{sol/KM}) the affine solution to the equations of motion. 

%------------------------------------------
\subsection{The anti-de Sitter solution}
\label{Sec-Virasoro}

In principle we could consider the algebra (\ref{KM}) as our
definition for the gauge-fixed basic quantum commutator.  As we have
pointed out in Sec. \ref{Unitary}, the gauge field $\hat A$ is a real
$SU(2)$ current and therefore the modes $T^a_n$ satisfy the Hermitian
condition $(T^a_n)^\dagger=T^a_{-n}$.  We shall explore the quantum
properties of this general metric elsewhere. Our goal here
is to describe the space of solutions to general relativity with
anti-de Sitter boundary conditions. The metric which follows from the
general affine solution does not satisfy this requirement and thus we
need to impose further restrictions on it \cite{CHvD}.     

In the conventions displayed in (\ref{conventions}), the solution
(\ref{sol/KM}) for $A_w$ can be written as, 
\begin{equation}
A_w =  {i\over 2} b^{-1} \left(\begin{array}{cc}
                    \hat A^3  &  \hat A^+ \\
                   \hat A^-  &   - \hat A^3
                   \end{array} \right) b
\label{A}
\end{equation}
with $\hat A^\pm = \hat A^1  \pm i \hat A^2$ and $b$ is defined in
(\ref{b}). We remind that the other components of the solution are
$A_{\bar w}=0$ and $A_\rho=iJ_3$, and similar expressions hold for the
anti-holomorphic field.   

The extra boundary conditions follow from looking at the form
of the gauge field associated to Euclidean anti-de Sitter space which
satisfies
\begin{equation}
\hat A^+ = -2, \ \ \ \ \hat A^3 = 0.
\label{red}
\end{equation}
The anti-holomorphic field satisfies $\hat A^-=-2,\hat A^3=0$. These
conditions
were first discussed in \cite{Polyakov90}, and their relation to
anti-de Sitter spaces was realised in \cite{CHvD}. In the $WZW$
approach, they can be incorporated in the action by adding Lagrange
multipliers. This leads to a gauged $WZW$ action whose conformal
generators follow from the GKO coset construction.

We shall impose (\ref{red}) as part of the boundary data. Since
(\ref{sol/KM}) solves the equations of motion for arbitrary values of
$\hat A^+,\hat A^-$ and $\hat A^3$, the gauge field will still be a
solution after imposing (\ref{red}).  This means that among the three
components of the gauge field $\hat A^a$ only one component, $\hat
A^-$, remains as an arbitrary function.  It is convenient to rename
this function as, 
\begin{equation}
L(w) = - {k \over 2}  \hat A^-(w) 
\label{L}
\end{equation}
and thus the solution we are interested in has the form,
\begin{equation}
A_w(L) =  -i b^{-1} \left(\begin{array}{cc}
        0  &  1  \\
     (1/k) L(w)  &  0
                   \end{array} \right)b,
\label{A/red}
\end{equation}
where $L(w)$ is an arbitrary function of $w$, and $b$ is given in
(\ref{b}). As we shall see, this solution is appropriated to anti-de
Sitter spacetimes. 

The boundary group leaving (\ref{red}) invariant is no longer the
Kac-Moody algebra (\ref{KM}) because that algebra does not preserve
(\ref{red}). Let us find the group of transformations leaving
(\ref{red}) invariant. First, we look for those gauge transformations
$\delta A = D\lambda$ which preserve the form (\ref{A/red}) changing
only the values of $L$.  These transformations are generated by
parameters of the form \cite{Polyakov90}, 
\begin{equation}
\lambda = 
b^{-1}\left( \begin{array}{cc}   
             (i/2) \partial \varepsilon    &  \varepsilon   \\
  (1/k)\varepsilon L + (1/2)\partial^2\varepsilon   & -(i/2) 
                                                   \partial\varepsilon     
\end{array} \right)b
\label{lambda/res}
\end{equation}
where $\varepsilon=\varepsilon(w)$ is an arbitrary function of $w$,
and $b$ is given in (\ref{b}). Acting with (\ref{lambda/res}) on $A_w$
we find, 
\begin{equation}
A_w(L) + D_w \lambda = A_w(L + \delta L),
\label{deltaA2}
\end{equation}
with
\begin{equation}
\delta L = i( \varepsilon \partial L + 2 \partial \varepsilon L +
{k\over 2} \partial^3 \varepsilon). 
\label{deltaL1}
\end{equation}
This shows that $L$ is a quasi-primary field of dimension two under
the residual group of transformations leaving (\ref{A/red}) invariant. 

The next step is to determine the algebra associated to these
transformations. This can be done by imposing the reduction conditions
(\ref{red}) in the algebra (\ref{KM}) and computing the induced
algebra. Geometrically speaking, given a Poisson bracket structure of
the form $[x^a,x^b] = J^{ab}(x^a)$ ($J^{ab}$ invertible) one defines
the symplectic form $\sigma_{ab}$ as the inverse of $J^{ab}$. The
antisymmetry and Jacobi identity satisfied by $J^{ab}$ imply that
$\sigma_{ab}$ is a closed 2-form. Now, let $\chi_\alpha(x^a) = 0$ a
set of constraints on phase space such that
$C_{\alpha\beta}:=[\chi_\alpha,\chi_\beta] $ is invertible. The
surface defined by $\chi_\alpha(x^a) = 0$ will be called $\Sigma$. Let
$\sigma^*$ the pull-back of $\sigma$ into $\Sigma$. It follows that
the induced Poisson bracket structure on $\Sigma$ is simply the
inverse of $\sigma^*$ (the invertibility of $\sigma^*$ is guaranteed
by the invertibility of $C_{\alpha\beta}$).  See, for example,
\cite{HT-book} (chapter 2) for more details on this construction and
in particular its relation with the Dirac bracket.            

In terms of the modes $T^a_n$ defined in (\ref{T}), conditions
(\ref{red}) read 
\begin{equation}
T^+_n=-k \delta^0_n, \ \ \ \ T^3_n=0
\label{red/mode}
\end{equation}
we then need to consider the matrix (evaluated on the surface
(\ref{red/mode})), 
\begin{equation}
C := \left( 
\begin{array}{cc}
          [T^+_n,T^+_m]  & [T^+_n,T^3_m]   \\
		 ~[T^3_n,T^+_m]  & [T^3_n,T^3_m]
\end{array} 
\right) = 
\left( 
\begin{array}{cc}
               0         &  k\delta_{n+m}  \\
		  -k\delta_{n+m} & (kn/2) \delta_{n+m}
\end{array} 
\right)
\end{equation}
which is indeed invertible. We remind here the form of the algebra
(\ref{KM}) in the basis $\{T^\pm = T^1 \pm i T^2$, $T^3_n\}$,  
\begin{eqnarray}
~[T^+_n,T^-_m] &=& 2T^3_{n+m} + nk \delta_{n+m}, \\
~[T^3_n,T^\pm_m] &=& \pm T^\pm_{n+m}, \\
~[T^3_n,T^3_m] &=&  {kn \over 2} \delta_{n+m}.
\end{eqnarray}

By an straightforward application of the
method explained above, the induced Poisson structure $[\ ,\ ]^*$ (or
Dirac bracket) acting on the surface (\ref{red}) can be written in
terms of the original bracket $[\ ,\ ]$ as (sum over $n\in Z$ is
assumed),
\begin{eqnarray}
[a,b]^* = [a,b] + {n\over 2k} [a,T^+_n][T^+_{-n},b]+ {1  \over k} 
[a,T^+_n][T^3_{-n},b] -{1 \over k} [a,T^3_n][T^+_{-n},b].
\label{Dirac}
\end{eqnarray}
This bracket, by definition, satisfies $[a,T^3_n]^*=0=[a,T^+_n]^*$ for
any function $a$, on the surface (\ref{red/mode}).  Now we compute the
algebra of the remaining component $T^-_n$. As before, it is
convenient to define $L_n= -T^-_n$. The $L_n$'s are then the Fourier
modes of the function $L$ defined in (\ref{L}), 
\begin{equation}
L(w) = \sum_{n\in Z} L_n e^{inw}
\label{LF}
\end{equation}
and satisfy the Virasoro algebra,
\begin{equation}
[L_n,L_m]^* = (n-m)L_{n+m} - {k \over 2} n^3 \delta_{n+m}
\label{Virasoro}
\end{equation}
with a central charge 
\begin{equation}
c=-6k. 
\label{ck}
\end{equation}
Note that since $k$ is negative (see (\ref{k})), the
central charge in (\ref{Virasoro}) is positive and thus highest weight
unitary representations exist.     

The form of the central term in (\ref{Virasoro}) is not the standard
one. One could shift $L_0$ in order to find the usual $n(n^2-1)$ term.
However, for the black hole whose exact isometries are $SO(2)\times
SO(2)$  (due to the identifications) it is more natural to leave the
central term as in (\ref{Virasoro}).  This is also natural from the
point of view of supergravity since the vacuum black
hole Killing spinors are periodic \cite{Coussaert-H}.   

The space of solutions (\ref{A/red}) is invariant under conformal
transformations generated by $L(w)$. Note that in the Chern-Simons
formulation of  three-dimensional gravity the Chern-Simons  coupling
$k$ was related to Newton's constant $G$ as $k=-l/4G$ (see (\ref{k})).
This means that the central charge in the Virasoro algebra $c=-6k$
coincides with the Brown-Henneaux \cite{BH} central charge $c=3l/2G$.
This is not a coincidence \cite{CHvD}. As we shall see, the above
conformal algebra represents exactly the Brown-Henneaux conformal
symmetry of three-dimensional adS gravity. The relation between the
reduction conditions (\ref{red}) and the conformal symmetry found in
\cite{BH} was stablished in \cite{CHvD}.  A previous calculation of
the central charge using the Chern-Simons formulation of
three-dimensional gravity and a twisted Sugawara construction was
presented in \cite{B}. 

What we have done for the holomorphic sector can be repeated for the
anti-holomorphic sector. The reduction conditions in this case read
$\bar T^-_n=-k\delta_n^0$ and $\bar T^3_n=0$. Since the affine
$SU(2)_k$ algebra is invariant under the change $T^+_n \leftrightarrow
T^-_n$, $T^3 \rightarrow -T^3_n$, one finds the same induced algebra.
The anti-de Sitter solution for the anti-holomorphic part reads 
\begin{equation}
\bar A_{\bar w}(\bar L) = 
  -i b \left( \begin{array}{cc} 0 & (1/k)\bar L(\bar w) \\
           1     &  0   \end{array} \right)b^{-1}, 
\label{sol/Vir/bar}
\end{equation}
plus $\bar A_w=0$ and $\bar A_\rho = -iJ_3$.  The residual gauge
transformations are  
\begin{equation}
\bar \lambda =  
b\left( \begin{array}{cc}   
-(i/2)\partial\bar\varepsilon & (1/k)\bar\varepsilon\bar
L+(1/2)\partial^2\bar\epsilon \\
\bar\varepsilon &(i/2)\partial\bar \varepsilon     
\end{array} \right)b^{-1} , 
\label{lambda/res/bar}
\end{equation}
where $\bar\varepsilon=\bar\varepsilon(\bar w)$ is an arbitrary
function of $\bar w$. Again $\bar L$ is a Virasoro operator and the
central charge is $c=-6k$.  Note that the effective coupling in the
anti-holomorphic Chern-Simons theory is $-k$ because the Chern-Simons
action $I[\bar A]$ has a
minus sign in front. However, the non-zero current is now $\bar
A_{\bar w}$ instead of $\bar A_w$. The change $k\rightarrow -k$ is
compensated by $w\rightarrow \bar w$ and we find the same Virasoro
algebra with the same central charge in both cases.    

We shall call the gauge fields (\ref{A/red}) and (\ref{sol/Vir/bar})
the anti-de Sitter solution of the equations of motion because they
are appropriated to anti-de Sitter spacetimes.  

%-------------------------------------
\subsection{The $WZW$ and Liouville actions} 
\label{WZW}

We have displayed in this section two solutions to the Chern-Simons
equations of motion with an affine (Sec. \ref{Sec-KM}) and Virasoro
(Sec. \ref{Sec-Virasoro}) symmetries.  We have also argued that, at
least classically, the
corresponding algebras can be interpreted as basic Poisson brackets
acting on the corresponding phase spaces.  A natural and powerful way
to justify this point is by studying the induced theories at the
boundary for the corresponding boundary conditions. It is known
\cite{Moore-S} that under  the boundary condition (\ref{bcond}), the
Chern-Simons action reduces to a $WZW$ action at the boundary whose
basic Poisson bracket, first calculated in \cite{Witten84}, is the
affine algebra (\ref{KM}).  Reducing the affine 
solution via (\ref{red}) then gives (\ref{Virasoro}). At this point, a
natural question to ask is what is the boundary action (analogue to
the $WZW$ action) which would give rise directly to the Virasoro
algebra (\ref{Virasoro}) as its basic Poisson bracket. We do not know
the answer to this question.  However, an alternative route can be
taken. It was shown in \cite{CHvD} that the two chiral $WZW$ actions
arising in Chern-Simons gravity can be combined into a single
non-Chiral action via $g=g^{-1}_1 g_2$. Furthermore, the
reduction conditions (\ref{red}) applied to the non-Chiral theory 
lead to a Liouville action \cite{S} which has the 
expected conformal symmetry with a central charge equal to $c=-6k$.
The solutions of three-dimensional gravity can be classified in
terms of the solutions of Liouville theory. One finds that the
different monodromy conditions (elliptic, parabolic and hyperbolic)
led to the three classes of solutions: conical singularities,
extreme, and black holes\cite{Martinec98}.  See \cite{Navarro} for a
direct relation between the Liouville energy momentum tensor and the
anti-de Sitter boundary conditions, without using the Chern-Simons
formalism.  The Liouville field has also appeared in
\cite{Deser-Jackiw} in the study of solutions to the 2+1 classical
equations with a positive cosmological constant.   

The Liouville action is certainly a good candidate to describe the
dynamics of 2+1 gravity with anti-de Sitter boundary conditions.
However, it should be kept in mind that its derivation from the $WZW$
model is not unique and a better control on some global issues is
necessary. First, merging the two Chiral $WZW$ actions into a single
one through $g=g^{-1}_1 g_2$ is not unique because $g$ is invariant
under $g_1 \rightarrow Ag_1$, $g_2 \rightarrow Ag_2$ with $A$
arbitrary. Second, the Liouville action arises in terms of a Gauss
decomposition of the group element which is not global. For these
reasons we have not committed ourselves to any particular form for the
boundary action but instead we have treated the basic Poisson bracket
algebras as the starting point for quantization.

%-------------------------------------------------------
%-------------------------------------------------------
\section{A quantum spacetime}
\label{Metric}

We have found in the last section a general solution for the
classical equations of motion with prescribed boundary conditions.
We have also found the induced Poisson bracket structure acting on
the gauge-fixed dynamical functions. Our aim in this section is to
quantise those spaces and apply the results to three-dimensional
gravity.   

However, an important warning is necessary here. The space of
solutions that we have displayed (affine and anti-de Sitter solutions)
are explicitly not coordinate invariant. This means that we could have
chosen other coordinates to describe that space and it is not
guaranteed that the corresponding quantum versions would be
equivalent. For example, we know that a full quantization of the
Chern-Simons action with compact groups induces a shift in the
coupling constant \cite{Witten89,Moore-S,Labastida-R} which is not
seen in our gauge-fixed approach (although it is suggested by the
Sugawara form of the conformal generators; see \cite{B} for a
construction of the Virasoro generators using a twisted Sugawara
operator).  We shall not attack this problem here in the belief that
at least in the large $k$ limit our results can be trusted.

It is worth stressing here that the fact that we have found
non-Abelian Poisson structures (Virasoro and Kac-Moody algebras) is a
consequence of the self-interacting character of gravity. The
non-Abelian pieces in those algebras are measured by the coupling $k$
which in the semiclassical limit, when gravity becomes linearized,
goes to infinity.    

\subsection{The metric} 

The general solution to Einstein equations in three dimensions
with anti-de Sitter boundary conditions can be found using the results
of last section together with the correspondence between metrics
$g_{\mu\nu}$ and connections $A^a_\mu, \bar A^b_\nu$ discovered in
\cite{Achucarro-T}. Given the connections $A^a_\mu, \bar A^a_\mu$, one
constructs the Lorentz vector $e^a_\mu = (l/2i)( A^a_\mu - \bar
A^a_\mu)$ and then the metric $g_{\mu\nu} = e^a_\mu e^b_\nu
\delta_{ab}$. It follows from the analysis of \cite{Achucarro-T} that
if $A^a_\mu$ and $\bar A^a_\mu$ satisfy the Chern-Simons equations of
motion, then $g_{\mu\nu}$ satisfies the three-dimensional
gravitational equations.

For our calculations it will be useful to define the matrix,
\begin{equation}
e_\mu = {l\over 2i} ( A_\mu - \bar A_\mu)
\label{e/AA}
\end{equation}
where $A = A^a J_a$ and $\bar A = \bar A^a J_a$. The spacetime metric
is then given by
\begin{equation}
g_{\mu\nu} = -2 \mbox{Tr} (e_\mu e_\nu).
\label{g/ee}
\end{equation}
The conventions for the $J_a$ matrices are displayed in
(\ref{conventions}).

The affine solution (\ref{sol/KM}) (and its anti-holomorphic part)
is certainly a very general and
interesting solution for the Chern-Simons equations of motion,
however, the induced metric does not satisfy the anti-de Sitter
boundary conditions prescribed in \cite{BH}. This is the reason that
we have also considered the reduction of (\ref{sol/KM}) via
(\ref{red}) which leaves an asymptotically anti-de Sitter spacetime,
with a conformal symmetry. 

Consider the anti-de Sitter solutions (\ref{A/red}) and
(\ref{sol/Vir/bar}) for the Chern-Simons equations of motion and let
us compute the associated metric. Using (\ref{e/AA}) we compute the
components of the triad,  
\begin{eqnarray}
   e_w    &=& -{l\over 2} \left( \begin{array}{cc}  
                           0  &  e^\rho  \\
                          e^{-\rho} L/k  &  0   
	    \end{array} \right),  \label{ez} \\
e_{\bar w} &=& {l\over 2}   \left( \begin{array}{cc} 
                           0  & e^{-\rho} \bar L/k    \\
                       e^\rho &  0 
		\end{array} \right),  \label{ebarz} \\
e_\rho &=& lJ_3,	\label{erho}	
\end{eqnarray}
and then using (\ref{g/ee}) we find the metric,
\begin{equation}
 ds^2 = 4Gl( L dw^2 + \bar L d\bar w^2) +
\left(l^2 e^{2\rho} + 16 G^2 L \bar L e^{-2\rho} \right)dw d\bar w +
l^2 d\rho^2 .
\label{dsLLq}
\end{equation}
Here we have used the value of $k$ given in (\ref{k}). This is the
metric that was displayed in Sec. \ref{Brief}. We can
now go through the properties listed in that section and
check their validity. 

First, by construction (\ref{dsLLq}) solves the Einstein equations
because the corresponding gauge fields solve the Chern-Simons
equations. This, of course, can be proved explicitly by checking that
(\ref{dsLLq}) has constant curvature. Since $L(w)$ and $\bar L(\bar
w)$ are arbitrary functions, the metric (\ref{dsLLq}) provides an
infinite number of solutions to Einstein equations with a negative
cosmological constant in three dimensions. These solutions represent
different physical states because two metrics with different values of
$L,\bar L$ are related by a global diffeomorphism (global
diffeomorphism here is what was called ``improper" in
\cite{Benguria-CT}). At this point it is necessary to prove that the
notion of global diffeomorphism has an analogue within general
relativity. Since (\ref{dsLLq}) has constant curvature, there exists a
change of coordinates mapping (\ref{dsLLq}) into the anti-de Sitter
metric (\ref{adS}). The question is whether that change of coordinates
is generated by one of the constraints of general relativity or not.  

In complete analogy with the gauge case, we define global
diffeomorphisms as coordinate transformations which are not generated
by the constraints of general relativity\cite{Regge-CT}. Rather,
global diffeomorphisms are generated by non-zero quantities which
enter as boundary terms in their canonical generators. This has been
analysed in detail in \cite{BH} from where we conclude that two
metrics of the form (\ref{dsLLq}), which only differ of the values of
$L$ and $\bar L$, are connected by a global diffeomorphism. The
functions $L$ and $\bar L$ then represent physical degrees of freedom
from the gravitational point of view.    

%-----------
\subsection{Diffeomorphisms in Chern-Simons gravity}
\label{Diff/CS-GR}

It is interesting and instructive to prove explicitly that there
exists a change of coordinates $\{w,\bar w,\rho\} \rightarrow
\{w',\bar w',\rho'\}$ which preserve the form of (\ref{dsLLq})
changing only the values of $L$ and $\bar L$.  This is the goal of
this paragraph. To find these 
transformations we can either do it by brute force acting with Lie
derivatives on (\ref{dsLLq}), or by using the results of the last
sections making a dictionary between gauge transformations and
diffeomorphisms. We shall follow this last procedure.  

It is well known that, due to the flatness of the gauge field, in
Chern-Simons theory the diffeomorphism invariance is not an
independent symmetry. Indeed, a diffeomorphism along a vector field
$\xi^\mu$ can be  written as a gauge transformation with a parameter
$\lambda^a = A^a_\mu \xi^\mu$ \cite{Witten88}. The converse is, in
general, not true. However, in the case of Chern-Simons gravity where
the relevant group is $SL(2,C)$ and an invertible triad exists, one
can prove that all gauge transformations act on the metric via 
diffeomorphisms.

More explicitly, let $g_{\mu\nu}$ the metric associated to a
particular configuration $A^a_\mu,\bar A^b_\nu$ through (\ref{e/AA})
and
(\ref{g/ee}). Now, consider an arbitrary gauge transformation with
parameters $\lambda^a,\bar\lambda^a$ acting on $A^a,\bar A^a$. It
follows that the transformed metric (associated to the transformed
fields) is related to the original one by a diffeomorphism generated
by a vector field $\xi^\mu_{(\lambda,\bar\lambda)}$.  

To prove this statement, and find the explicit formula for
$\xi^\mu_{(\lambda,\bar\lambda)}$, consider the action of the gauge
group on the triad. From (\ref{e/AA}) we find that  under a gauge
transformation $\delta A_\mu = D_\mu \lambda$, $\delta \bar A_\mu =
\bar D_\mu \bar \lambda$ the triad changes according to, 
\begin{equation}
\delta e_\mu = {l\over 2i} D^{(w)}_\mu (\lambda-\bar\lambda) -{1\over
2} [e_\mu, \lambda + \bar \lambda].
\label{deltae}
\end{equation}
Here we have used that $A_\mu+\bar A_\mu = 2\omega_\mu $ where
$\omega_\mu = \omega^a_\mu J_a$ is the spin connection, and
$D^{(w)}_\mu$ denotes its associated covariant derivative. The second
term in (\ref{deltae}) is a Lorentz rotation of the triad which does
not change the metric.  We then concentrate on the first term.  Let us
define the $SO(3)$ vector $\rho^a$ and its associated vector field
$\xi^\mu_{(\rho)}$ by,
\begin{eqnarray}
\rho^a  &=& {l\over 2i} (\lambda^a-\bar\lambda^a), \label{rho} \\
\xi^\mu &=& e^\mu_a \rho^a . \label{xi'}
\end{eqnarray}
We assume that $e^a_\mu$ is invertible then (\ref{xi'}) does make
sense. We now study how does the transformation,
\begin{equation}
\delta e^a_\mu = D^{(w)}_\mu \rho^a = D^{(w)}_\mu (e^a_\nu \xi^\nu),
\label{deltae2}
\end{equation}
change the metric.  Define the Christoffel symbols in the
standard way by $D^{(w)}_\mu e^a_\nu =  \Gamma^\sigma_{\nu\mu}
e^a_\sigma$ \footnote{The meaning of this definition can be uncovered
by writing it in the form $\Gamma^\sigma_{\mu\nu} = e_a^\sigma
\omega^a_{\ b\nu} e^b_\mu + e^\sigma_a e^a_{\mu,\nu}$. This is the
transformation law of a connection, $\omega^a_{\ b\nu} \rightarrow
\Gamma^\sigma_{\ \mu\nu}$, under the change of basis from the
coordinate basis $\partial_\mu$ to the orthonormal frame $\vec{v}_a$
described by the matrix $e^a_\mu$: $\partial_\mu = e^a_\mu
\vec{v}_a$.}. The transformation (\ref{deltae2}) becomes $\delta
e^a_\mu = e^a_\nu \xi^\nu_{\ ;\mu}$ where the semicolon denotes 
standard covariant derivative.  The action of this transformation
on the metric is, 
\begin{eqnarray}
\delta g_{\mu\nu}&=& \delta e^a_\mu e_{a\nu} + e^a_\mu \delta
e_{a\nu},\nonumber\\ 
           &=& (e^a_\sigma \xi^\sigma_{\ ;\mu}) e_{a\nu} + e^a_\mu
           (e_{a\sigma} \xi^\sigma_{\ ;\nu}), \nonumber \\
		   &=& \xi_{\mu;\nu} + \xi_{\nu;\mu},
\label{deltag}
\end{eqnarray} 
where in the last line we have used the definition of the metric
tensor and the identity $g_{\mu\nu;\sigma} =0$.  Thus, a
transformation in the triad of the form (\ref{deltae2}) is indeed seen
in the metric as a diffeomorphism. Since the gauge transformations
acting on $A,\bar A$ produce (up to a Lorentz rotation) a
transformation of the form (\ref{deltae2}) with $\rho^a$ given in
(\ref{rho}), we conclude that the gauge group acts on the metric via a
diffeomorphism with a parameter defined in (\ref{xi'}).  

Now we apply this result to the particular case of the residual gauge
transformations (\ref{lambda/res}) and (\ref{lambda/res/bar}). The
residual vector field $\xi^\mu_{(\varepsilon,\bar\varepsilon)}$
associated to those transformations is computed directly from
(\ref{lambda/res}) and (\ref{lambda/res/bar}) plus (\ref{rho}) and
(\ref{xi'}). The formulae for the triad are given in
(\ref{ez})-(\ref{erho}).

It should be clear from the above analysis that
$\xi^\mu_{\varepsilon,\bar\varepsilon}$  generates a residual symmetry
of the metric (\ref{dsLLq}). This can be summarised in the following
table.  $A_\mu(L)$ and $\bar A^a_\nu(\bar L)$ represent the residual
connections (\ref{A/red}) and (\ref{sol/Vir/bar}), 
and $g_{\mu\nu}(L,\bar L)$ the associated metric (\ref{dsLLq}). Under
the residual gauge transformations (\ref{lambda/res}) and
(\ref{lambda/res/bar}) the gauge field and metric transform according
to:   \\

\fbox{
$\begin{array}{cccc}
 \begin{array}{c} A_\mu(L) \\ \bar A_\nu(\bar L) \end{array}
  \rightarrow &  
  \begin{array}{c}  A_\mu(L)+\delta A_\mu \\   \bar A_\nu(\bar
  L)+\delta \bar A_\nu  \end{array}  
  &  \begin{array}{c} = \\ = \end{array}   & 
  \begin{array}{c}  A_\mu(L+\delta L) \\ \bar A_\nu(\bar L + \delta
  \bar L) \end{array}  \\
  \Downarrow      &  \Downarrow     & &   \Downarrow   \\
  g_{\mu\nu}(L,\bar L)  \rightarrow & g_{\mu\nu}(L,\bar L) + 
  {\cal L}_{\xi_{(\varepsilon,\bar\varepsilon)}} g_{\mu\nu} &=& 
  g_{\mu\nu}(L+\delta L,\bar L + \delta \bar L) 
\end{array}$ 
}  
~\\
\\

\noindent where  $\delta L$ is given in (\ref{deltaL1}), and the same
expression holds for $\delta \bar L$.  The equalities in the first two
lines simply express the fact that the residual gauge transformations
leave the gauge field invariant changing only the values of $L$ and
$\bar L$. The third line contains non-trivial information. First, as
we discussed above, the metric associated to the transformed gauge
fields is related to the original metric via a diffeomorphism
parametrised with the residual vector field
$\xi^\mu_{(\varepsilon,\bar\varepsilon)}$. Then, since the transformed
metric can also be written in terms of $A(L+\delta L), \bar A(\bar
L+\delta \bar L)$, we find the last equality and conclude that the
vector field $\xi^\mu_{(\varepsilon,\bar\varepsilon)}$ generates a
residual diffeomorphism of the metric (\ref{dsLLq}). This analysis
exhibit the power of the Chern-Simons formalism.  To discover that the
metric (\ref{dsLLq}) has a residual conformal invariance --not only
asymptotically-- using Lie derivatives would have been extremely
complicated.      

Let us work out explicitly the case on which $\bar L=0$. The metric
(\ref{dsLLq}) reduces to the simple form 
\begin{equation}
 ds^2 = 4Gl L(w) dw^2 + l^2 e^{2\rho} dw d\bar w + l^2 d\rho^2. 
\label{ds'}
\end{equation}
We act on this metric with the holomorphic residual
transformation generated by (\ref{lambda/res}). The associated
residual vector $\xi^\mu=\delta x^\mu$ is computed from
(\ref{lambda/res}) and the triad (\ref{ez})-(\ref{erho}) with $\bar
L=0$. In the coordinates $\{w,\bar w,\rho\}$ it reads, 
\begin{equation}
\delta\rho=-{i\over 2} \partial \varepsilon, \ \ \ \ \delta w =
i\varepsilon, \ \ \ \ \delta \bar w = -{i \over 2}e^{-2\rho}
\partial^2
\varepsilon .  
\label{xi}
\end{equation}
Transforming the metric (\ref{ds'}) with this vector one finds the
same metric with $L$ replaced by $L'=L+\delta L$, and 
\begin{equation}
\delta L = i( \varepsilon \partial L + 2 \partial\varepsilon L -
{l\over 8G} \partial^3 \varepsilon ).
\label{deltaL2}
\end{equation}
Since $l/8G=c/12=-k/2$ with $c$ and $k$ given respectively in
(\ref{c}) and (\ref{k}), we find consistency with (\ref{deltaL1}), as
expected.  As a further check, we can now transform (\ref{ds'}) with
the anti-holomorphic residual transformation generated by
(\ref{lambda/res/bar}). Since $\bar L$ is a quasi-primary field, we
expect that this transformation will not preserve $\bar L=0$ and thus
the metric (\ref{ds'}) will be transformed into (\ref{dsLLq}) with
$\bar L = \delta \bar L$. Indeed, from (\ref{lambda/res/bar}) and
(\ref{ez})-(\ref{erho}) we find the associated transformation, 
\begin{equation}
\delta\rho=-{i\over 2} \bar\partial \bar\varepsilon, \ \ \ \ 
\delta w = -{i \over 2}e^{-2\rho}\bar\partial^2\bar\varepsilon,\ \ \ \ 
\delta \bar w = i\varepsilon  + {2i G L\over l} e^{-4\rho}
\bar\partial^2 \bar\varepsilon .     
\label{xi/bar}
\end{equation}
We act on (\ref{ds'}) with this vector and find a metric of the form
(\ref{dsLLq}) with $\bar L = (-il/8G)\bar\partial^3\bar\varepsilon $.
This is exactly the right transformation, in accordance with
(\ref{deltaL2}) applied to the anti-holomorphic field $\bar L(\bar
w)$.    

As a last example of the conformal residual symmetries of
(\ref{dsLLq}) we mention here the case of the finite (holomorphic) 
exponential map. To simplify the notation we set here $4G=1$ and
$l=1$. Let us make the finite change of coordinates on (\ref{ds'})
$\{w,\bar w,\rho\} \rightarrow \{z,\bar w',\rho'\}$ defined by
\begin{equation}
z = e^{-iw}, \quad \bar w' = \bar w + (i/2) e^{-2\rho}, \quad 
e^{2\rho'} = ie^{2\rho+i w}
\label{exp/map}
\end{equation}
This transformation maps the cylinder $w$ into the plane $z$, but
leaves $\bar w$ in the cylinder.  Note that here we are using
explicitly the independence of $w$ and $\bar w$. We act on the metric
(\ref{ds'}) with this change of coordinates and find again the same
metric with $L(w)$ replaced by $T(z) = -z^{-2} (L(w)+ 1/2)$. The shift
$1/2=6/12$, of course, corresponds to the Schwarztian derivative of
the map (\ref{exp/map}). (In the conventions $4G =1$ and $l=1$ the
central charge (\ref{c}) is equal to 6.)     

%------------------
\subsection{Black holes}
\label{Sec-BTZ}

As we have mentioned above, the metric (\ref{dsLLq}) reduces
to a three-dimensional black hole \cite{BTZ,BHTZ} when $L=L_0$ and
$\bar L=\bar L_0$ are constants. This can be proved as follows. First,
we define the constants $M$,$J$ and $r_\pm$ by  
\begin{eqnarray}
 L_0 + \bar L_0  &= Ml =&  {r_+^2 + r_-^2 \over 8Gl}, \label{M}\\ 
 L_0 - \bar L_0  &= J =&  {2 r_+ r_- \over 8Gl}. \label{J}
\end{eqnarray}
Next we define the real coordinates $\varphi,t$ and $r$ by, 
\begin{equation}
w = \varphi + i t,
\label{w0q}
\end{equation}
and 
\begin{equation}
r^2 = r_+^2 \cosh^2(\rho-\rho_0) - r_-^2 \sinh^2(\rho-\rho_0).
\label{r}
\end{equation}
The constant $\rho_0$ is given by $e^{2\rho_0}= (r_+^2-r_-^2)/(4l^2)$.
This radial definition has the property $l^2 d\rho^2 = N^{-2}
dr^2$ where $N^2$ is the lapse function appearing in the black hole
metric (\ref{BTZ}). The constants $r_\pm$ are the solutions of the
equation $N^2(r_\pm)=0$.  After a long but direct calculation one can
prove that the metric (\ref{dsLLq}), in the coordinates
$\{t,r,\varphi\}$, is exactly equal to the metric (\ref{BTZ}) with
mass $M$ and angular momentum $J$. 

Since we are working in the Euclidean sector, the coordinate $t$
appearing in (\ref{w0q}) and (\ref{BTZ}) is periodic, $0\leq t
<\beta$,
with $\beta = 2\pi l^2 r_+/(r_+^2-r_-^2)$.  In order to fix the period
of the time coordinate to be independent of the black hole parameters
(and thus fix the complex structure of the torus), one can define
$z=\varphi + \tau x^0$ with $0\leq x^0<2\pi$ and $\tau = i\beta/2\pi$.
Since $\varphi$ and $x^0$ are periodic, the complex coordinate $z$ is
defined on a torus,
\begin{equation}
z \sim z + 2\pi n + 2\pi \tau m , \ \ \ \ \  n,m \in Z. 
\label{torus}
\end{equation}
with $\tau$ its modular parameter. Introducing $\tau$ is particularly
convenient when studying modular invariance on the black hole
manifold\cite{BBO,Maldacena-S}. We
shall not deal with this issue here, so we use (\ref{w0q}). 

%------------------------------
\subsection{The quantum space of metrics. State counting}
\label{Quantum-met}

We have described in the last paragraph a set of metrics parametrized
by two functions whose induced Poisson bracket yield the Virasoro
algebra with a non-zero central charge. We shall now promote the
algebra (\ref{Virasoro}) to be a quantum algebra and study the
properties of the associated quantum metric. 

The unitary representations of the Virasoro algebra for a given
positive central charge $c$ are parametrized by a single real positive
number $h$. In the semiclassical limit with $-k$ large, the Virasoro
central charge $c=-6k$ is then large. Under these conditions, there
exists one unitary representation for each conformal dimension $h$. We
start with the vacuum state $|h>$ satisfying $L_0|h> = h|h>$ and $L_n
|h>=0$ ($n>0$). The excited states are constructed with the negative
modes $L_{-n}$ acting on $|h>$. The full representation, for a given
$h$, is spanned by the vectors $|n_1,...,n_r;h> := L_{-n_1} \cdots
L_{-n_r} |h>$ with $r=1,2,...$. The same construction has to be
repeated for the other Virasoro algebra $\bar L_n$. 

In standard conformal field theory, the values of $h$ are not
arbitrary. They are equal to the conformal dimensions of the primary
fields $\phi_h$ of the theory. The state $|h>$ is created by $\phi_h$ 
via $|h> = \phi_h(0) |0>$ where $|0>$ is the true conformal vacuum.
In our situation, we do not have a field theory at the boundary
(of course it could be Liouville theory, see Sec. \ref{WZW} for a
discussion on this point) but only the Virasoro algebra. The usual
state-operator map will then be missing until we decide which is right
conformal field theory.     

Once we promote the modes $L_n$ and $\bar L_m$ to be operators acting
on Fock space, the metric (\ref{dsLLq}) becomes an operator that we
shall denote by $ d\hat s^2$. Note that since the metric (\ref{dsLLq})
is an algebraic function in $L$ and $\bar L$ which does not involve
products of non-commuting operators, $ d\hat s^2$ is well defined in
the operator sense. 

We can now define a natural map from Fock space to the
space of classical solutions.  For each state $|\Psi>$ in Fock space,
we associate a classical solution to the equations of motion given by, 
\begin{equation}
ds^2_\psi = <\psi| d\hat s^2 |\psi>
\label{clas}
\end{equation}
where $ds^2_\psi$ is the metric (\ref{dsLLq}) with $L=<\psi| \hat L
|\psi>$ and  $\bar L=<\psi| \hat{ \bar L} |\psi>$.  Since
(\ref{dsLLq}) is a solution for arbitrary values of $L$ and $\bar L$,
the metric (\ref{clas}) is a solution for any state $|\psi>$. More
interesting, the full set of solutions (\ref{dsLLq}) can be generated
by the above map.   

According to (\ref{clas}), every state $|\psi>$ induces a unique
classical solution. The converse is not true. For a given
classical solution there may be many associated states.  In
particular, there are many states associated to a given black hole of
mass $M$ and angular momentum $J$. As explained before, the metric
(\ref{dsLLq}) gives rise to a black hole when $L$ and $\bar L$ are
constants and related to $M$ and $J$ by (\ref{M}) and (\ref{J}). 
Let $|M,J;\lambda>$ the set of states in the Hilbert space such that
they satisfy 
\begin{eqnarray}
<M,J;\lambda|(L_n+\bar L_n)|M,J;\lambda> &=& lM\, \delta^0_n,
\nonumber \\ 
<M,J;\lambda|(L_n - \bar L_n)|M,J;\lambda> &=& J \, \delta^0_n ,
\label{qbh}
\end{eqnarray}
for all $\lambda = 1,2,3,...,\rho(M,J)$. These states generate through
(\ref{clas}) a black hole of mass $M$ and angular momentum $J$.  We
can then formulate the problem of black hole degeneracy as whether the
logarithm of the number of these states, $\ln \rho(M,J)$, is equal to
the Bekenstein-Hawking entropy of the corresponding black hole of mass
$M$ and angular momentum $J$ or not. The answer to this question
depends on the structure of the Hilbert space.

Let us first work under the assumption that the Virasoro algebra is
the basic quantum commutator of the theory. In this case, the 
counting is very simple.  The states $|n_1,...,n_r;h>$, properly
normalized, precisely have the property, 
\begin{equation}
<n_1,...,n_r;h|L_n|n_1,...,n_r;h> = L_0 \delta^0_n  \ \ \ \
\mbox{with}
\ \ \ \ L_0 = h + \sum_{i=1}^r n_i.
\end{equation}
The number of these states, $\rho(L_0,\bar L_0)$, is then equal to
number of ways that one can write an integer as a sum of integers. For
large values of $L_0$ and $\bar L_0$ this number is approximated by
the well-known Ramanujan formula, 
\begin{equation}
\rho_{c'}(L_0,\bar L_0)=e^{2\pi \sqrt{c'L_0/6} + 2\pi \sqrt{c'\bar 
L_0/6}},
\label{Ramanujan}
\end{equation}
with $c'=1$.  Unfortunately, this naive counting does not
give the right result. Inserting $lM=L_0+\bar L_0$ and $J=L_0-\bar
L_0$ in (\ref{Ramanujan}) gives an entropy equal to $S = c^{-1/2}
A/4G$, where $c$ is the central charge (\ref{c}) and $A=2\pi r_+$ is
the perimeter of the horizon.  (The relation between the different
parameters is given in (\ref{M}) and (\ref{J}).)  The prefactor
$c^{-1/2}$ shows that our naive procedure is not yet correct because
we would expect the degeneracy of states to be equal to the
Bekenstein-Hawking value. 

If we do not regard (\ref{Virasoro}) as the basic algebra but only as
representing the symmetry algebra of some underlying conformal field
theory, then an elegant and striking way to relate (\ref{Virasoro})
with the correct Bekenstein-Hawking entropy is available 
\cite{Strominger97}. Suppose that the algebra (\ref{Virasoro})
represents the Virasoro algebra associated to some conformal field
theory with central charge $c$.  Suppose also that this CFT is
unitary, in the sense that $L_0,\bar L_0>-c/24$ (note that
we are using the Virasoro generators which vanish for the vacuum black
hole), and that the partition function, 
\begin{equation}
Z[\tau] = \mbox{Tr}\, e^{2\pi i \tau L_0 - 2\pi i \bar\tau \bar
L_0}, 
\label{Ztau}
\end{equation}
is modular invariant. This means 
\begin{equation}
Z[\tau'] = Z[\tau], \ \ \ \ \ \tau' = {a\tau +b \over c\tau + d},  
\label{mod}
\end{equation}
for any $a,b,c,d\in Z$ and $ad-bc=1$.  Then, it follows \cite{Cardy}
that that number of states with $L_0$ and $\bar L_0$ fixed is again
given by (\ref{Ramanujan}) but this time with $c'=c$.  The associated
entropy is then exactly equal to the Bekenstein-Hawking value $S=
A/4G$ with $A=2\pi r_+$ equal to the perimeter of the horizon
\cite{Strominger97}.  

As stressed in \cite{Carlip98-1}, this result is too beatiful to be
wrong. Even more, recently \cite{Carlip98-2,Solodukhin}, it has been
shown that under some boundary conditions at the horizon, similar
results can be applied to higher dimensions.  These are exiting
results which bring closer the long standing dream of a statistical
mechanical description for the Bekenstein-Hawking entropy. However,
there remains to find the conformal field theory responsible for the
degrees of freedom and, most importantly, to determine whether general
relativity is enough to describe that CFT, or other degrees of freedom
like string theory are necessary. 

An alternative route to get the right counting was suggested in
\cite{B4}. For integer values of the central
charge $c$, there is a natural way to add degrees of freedom to the
theory in such a way that the counting yields the right result. The
idea is that the Virasoro algebra (\ref{Virasoro}) can be regarded as
a sub-algebra of another Virasoro algebra, with central charge 1 and
generators $Q_n$, via the formula, 
\begin{equation}
L_n = {1\over c} Q_{cn}.
\label{LQ}
\end{equation}
See \cite{Borisov-HS} and references therein for a detailed
description of this embedding. (The formula (\ref{LQ}) has also
appeared in \cite{Toppan}.)  The number of states associated to the
representations of the operators $Q_n$ is again given by
(\ref{Ramanujan}) with $c'=1$ and $L_0$ replaced by $Q_0$. Since by
(\ref{LQ}) $Q_0=cL_0$, this yields the right result when using
(\ref{M}) and (\ref{J}). The main problem with this approach is that
we do not know how to relate the gravitational degrees of freedom to
the generators $Q_n$. Perhaps one should look for other boundary
conditions, generalising (\ref{red}), which may give other conformal
structures, generalising (\ref{Virasoro}). This issue is presently
under investigation.   

Whether the Virasoro operators are fundamental variables or not, this
will not change our quantum geometry picture. The microscopical origin
of the black hole degeneracy is associated to different states (living
in the correct CFT) which generate the same classical metric through
(\ref{clas}).

%--------------------------
\section{Final remarks}

Maldacena \cite{Maldacena} has conjectured a duality between large $N$
super-conformal field theory in four dimensions and Type IIB string
theory compactified on adS$_5 \times S_5$.  This relation has become
known as adS/CFT correspondence due to the relation between the
symmetry groups in each theory.  The result of Brown and Henneaux
\cite{BH} relating adS$_3$ and a conformal algebra in 1+1 dimensions
can also be regarded as an adS/CFT correspondence. Note however that
contrary to the higher dimensional case, this relation involves
only asymptotic adS space whose isometry group is infinite
dimensional. In \cite{Martinec98,Giveon98-,Henningson-Skenderis}  the
relation between these two aspects of the adS/CFT correspondence has
been explored.  

Finally, we would like to mention here a surprising motivation to
study three-dimensional gravity. It has been shown in
\cite{Sfetsos-Skenderis98} and \cite{Hyum97} (see also the recent
review \cite{Skenderis99})  that there exists duality
transformations relating five-dimensional black holes with
three-dimensional ones.  This means that everything we can learn 
about three-dimensional quantum gravity can be useful to higher
dimensional situations.

\section*{Acknowledgments} 

The author would like to thank H. Falomir, R.E. Gamboa Saravi and F.A.
Schaposnik for the invitation to the Buenos Aires' Meeting ``Trends in 
Theoretical Physics II".  Useful discussions with M.Asorey and F.
Falceto are acknowledged.  Financial support from CICYT (Spain) grant
AEN-97-1680, and the Spanish postdoctoral program of Ministerio de
Educaci\'on y Ciencia is also acknowledged.

\end{document}